\documentclass[aip,reprint]{revtex4-1}

\usepackage{graphicx}
\usepackage{dcolumn}
\usepackage{bm}

\begin{document}

\title{Hyperfine magnetic field in ferromagnetic graphite}

\author{Jair C. C. Freitas}
\affiliation{Departamento de F\'{\i}sica, Universidade Federal do Esp\'irito 
Santo, Vit\' oria, Brazil}
\email{jairccfreitas@yahoo.com.br}

\author{Wanderl\~a L. Scopel}
\affiliation{Departamento de F\'{\i}sica, Universidade Federal do 
Esp\'irito Santo, Vit\' oria, Brazil}
\affiliation{Departamento de Ci\^encias Exatas, Universidade 
Federal Fluminense, Volta Redonda, RJ, Brazil}
\email{wlscopel@gmail.com}

\author{Wendel S. Paz}
\affiliation{Departamento de F\'{\i}sica, Universidade Federal do Esp\'irito 
Santo, Vit\' oria, Brazil}

\author{Leandro V. Bernardes}
\affiliation{Department of Physics, Federal University of S\~ao Carlos, P.O. 
Box 676, 13565-905, S\~ao Carlos, SP, Brazil}

\author{Francisco E. Cunha-Filho}
\affiliation{Department of Physics, Federal University of S\~ao Carlos, P.O. 
Box 676, 13565-905, S\~ao Carlos, SP, Brazil}

\author{Carlos Speglich}
\affiliation{Department of Physics, Federal University of S\~ao Carlos, P.O. 
Box 676, 13565-905, S\~ao Carlos, SP, Brazil}

\author{Fernando M. Araújo-Moreira}

\affiliation{Department of Physics, Federal University of S\~ao Carlos, P.O. 
Box 676, 13565-905, S\~ao Carlos, SP, Brazil}

\author{Damjan Pelc}
\affiliation {Department of Physics, Faculty of Science, University of Zagreb, 
Bijenička 32, HR-10000, Zagreb, Croatia}

\author{Tonči Cvitanić}
\affiliation {Department of Physics, Faculty of Science, University of Zagreb, 
Bijenička 32, HR-10000, Zagreb, Croatia}

\author{Miroslav Požek}
\affiliation {Department of Physics, Faculty of Science, University of Zagreb, 
Bijenička 32, HR-10000, Zagreb, Croatia}

\email{jairccfreitas@yahoo.com.br}

\date{\today}

\begin{abstract}
Information on atomic-scale features is required for a better understanding of 
the mechanisms leading to magnetism in non-metallic, carbon-based materials. 
This work reports a direct evaluation of the hyperfine magnetic field produced 
at $^{13}$C nuclei in ferromagnetic graphite by nuclear magnetic resonance 
(NMR). The experimental investigation was made possible by the results of 
first-principles calculations carried out in model systems, including graphene 
sheets with atomic vacancies and graphite nanoribbons with edge sites partially 
passivated by oxygen. A similar range of maximum hyperfine magnetic field values 
(18-21 T) was found for all systems, setting the frequency span to be 
investigated in the NMR experiments; accordingly, a significant $^{13}$C NMR 
signal was detected close to this range without any external applied magnetic 
field in ferromagnetic graphite.
 
 \end{abstract}
 

\maketitle

The occurrence of magnetism in carbon materials has been the subject of many 
investigations and some controversy along the past two decades. An enormous 
interest in the possibility of producing magnetic materials predominantly 
consisting of carbon (perhaps with other light elements such as hydrogen and 
oxygen, but free from metallic elements) exists mainly due to the potential for 
applications of biocompatible magnetic materials to be used in drug delivery and 
magnetic resonance imaging, among others. Moreover, the design of graphene-based 
spintronics devices would greatly benefit from the definitive establishment and 
a deep understanding of the mechanisms leading to magnetism in carbon materials 
\cite{yazyev,recher}. Recent experimental evidences of magnetic properties 
(with reports of ferromagnetic order in some cases) of carbon-based materials 
include irradiated graphite, nanocarbons, oxygen-containing carbons and point 
defects in graphene \cite{yazyev,makarova,nair,silva,ugeda,mombru}. From the 
theoretical point of view, magnetism in graphene and related 
materials has been universally associated with the occurrence of defects such as 
atomic vacancies, chemisorbed species (such as fluorine, hydrogen and oxygen) 
and edge sites \cite{yazyev,makarova,silva,ugeda,paz}. In spite of this, there 
is still some skepticism about the possibility of intrinsic magnetic effects in 
carbon-based materials, due to the ubiquitously questioned presence in 
experimentally-produced samples of minor amounts of iron or other metallic 
impurities that could be the actual source of magnetism \cite{yazyev,makarova}. 

If intrinsic magnetism is indeed a feature of carbon-based materials, then it 
would be possible in principle to look for the effects due to atomic magnetic 
moments on the respective atomic nuclei, i.e., a hyperfine interaction (HFI) 
should hopefully be detectable. In the case of systems possessing magnetic 
order, a strong hyperfine magnetic field (B$_{hf}$) at the atomic nuclei could 
be anticipated. Therefore, evidences of the HFI and measurements of B$_{hf}$ 
are highly desirable for a better understanding of the issue of magnetism in 
carbon-based materials, allowing the assessment of information on the source of 
magnetism from a local perspective. Besides its importance from a fundamental 
point of view, the study of the HFI is of high relevance for possible 
applications of graphene and related materials in spintronics and quantum 
information processing, since the HFI is considered the most important mechanism 
of electron spin decoherence in these systems \cite{yazyev2}. Moreover, the HFI 
plays a central role in many recent proposals of solid-state quantum computers 
\cite{kane,maurer}. There were some theoretical calculations regarding 
the HFI in graphene and related materials \cite{yazyev2,dora} and also a 
number of experiments were tried to detect B$_{hf}$ by using 
techniques such as muon spin spectroscopy (µSR) \cite{ricco} and perturbed 
angular distribution (PAD) \cite{mishra}. In none of these reports, however, 
any clue about the B$_{hf}$ value at $^{13}$C nuclei (Among the 
naturally occurring carbon nuclides, $^{13}$C is the only one with non-zero 
nuclear spin (I = 1/2) and therefore $^{13}$C nuclei are sensitive to the 
magnetic HFI in magnetic carbon-based materials) in a truly ferromagnetic 
carbon material was ever reported.

Paralleling the former reports about the HFI in magnetic metals such as iron, 
cobalt and nickel \cite{gossard,budnick}, the direct measurement of B$_{hf}$ 
at $^{13}$C nuclei should be possible in a ferromagnetic carbon-based material 
by means of zero-field nuclear magnetic resonance (NMR). However, contrary to 
the case of NMR experiments performed in diamagnetic or paramagnetic materials 
under strong applied magnetic fields, the main difficulty in zero-field NMR 
experiments is the lack of information about the possible frequency range where 
the resonance is to be found. We  thus decided to carry out a methodical 
investigation about the HFI in a ferromagnetic carbon-based material, searching 
for the resonance frequency and aiming to determine the value of B$_{hf}$ at 
$^{13}$C nuclei by means of zero-field NMR. The material chosen to perform the 
experiments was ferromagnetic graphite, produced by controlled oxidation of 
high-purity graphite \cite{request}. As previously described 
\cite{mombru,pardo,araujo}, this material can be produced in bulk 
quantities ($\sim$ 50 mg) and presents ferromagnetic order at room temperature 
and 
below. The maximum amounts of metallic impurities detected in this material are 
well below the limits required to account for its overall magnetization, 
pointing to a genuinely carbon-originated magnetism \cite{pardo}. The 
magnetic properties of this material are related to the defects introduced in 
the graphite lattice by the oxygen attack, in agreement with recent theoretical 
calculations performed in graphite nanoribbons with edges partially passivated 
by oxygen atoms \cite{silva}. The sample selected for the NMR experiments 
showed a well-defined hysteresis loop in a magnetization versus applied magnetic 
field measurement conducted at low temperature (1.8 K), with coercive field of 
ca. 500 Oe, clearly indicating its ferromagnetic character.

After choosing a suitable ferromagnetic carbon-based material, the next quest 
was to search for the zero-field NMR signal. The NMR experiments were conducted 
at low temperature (1.5 K), so as to maximize the sample magnetization and to 
minimize transverse relaxation effects in a two-pulse spin echo pulse sequence 
\cite{request,turov}. Without any clue about where to look for the 
resonance, it would be practically impossible to successfully pursue this task. 
Thus, a series of first-principles calculations based on the density functional 
theory (DFT) were carried out in model systems built to somewhat reproduce the 
local features of the structure of ferromagnetic graphite. The model systems 
chosen for the DFT calculations 
included graphene sheets with isolated or multiple single atomic vacancies as 
well as graphite nanoribbons with oxygen atoms adsorbed at the zigzag edge 
sites \cite{request}. These systems are known from previous reports 
\cite{yazyev,nair,silva,ugeda,paz,zhang} to give rise to magnetic moments 
localized at carbon atoms, with indications of a ferromagnetic ground state in 
some cases \cite{silva,zhang}, and thus they were considered good candidates for 
the initial calculations of B$_{hf}$ at $^{13}$C nuclei.

\begin{figure}[h!]
\begin{center}
\includegraphics[width= 8.5cm]{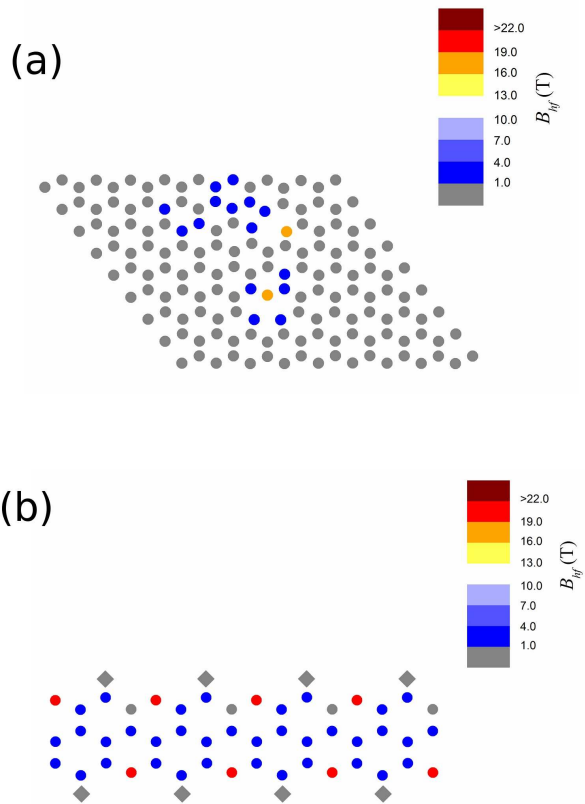}
\caption{Colormaps showing the distribution of calculated B$_{hf}$ values for 
system C (graphene sheet with two atomic vacancies ca. 11 \AA) (a) and along a 
single graphene layer in system E (graphite nanoribbon) (b). The circles 
represent carbon atoms and the diamonds in (b) represent oxygen atoms.}
\label{fig1}
\end{center}
\end{figure}

A summary of the results of the DFT calculations is presented in 
Table \ref{systems}. The first noticeable aspect of these results was that all 
B$_{hf}$ values fell into the same range, ~18-21 T, despite the different types 
of defects giving rise to magnetism in vacancy-containing graphene sheets and in 
oxygen-containing graphite nanoribbons. This is an indication that the B$_{hf}$ 
values here reported are indeed characteristic of carbon sites with localized 
magnetic moments in carbon-based systems. As it should be expected, the largest 
B$_{hf}$ values in each system were found at the sites also presenting the 
highest net spin densities and, thus, the largest atomic magnetic moments, as 
illustrated in Fig. \ref{fig1} for some of the studied systems.

\begin{table}[h!]
\begin{center}
\caption{Calculated hyperfine magnetic field (B$_{hf}$) in 
ferromagnetic carbon-based systems. The reported values correspond to the 
average taken over the sites where the largest magnetic moments were located in 
each system.}
\label{systems} 
\begin{tabular}{cc}
\hline\hline
System &B$_{hf}$(T) \\
\hline
Graphene sheet with one single vacancy\\ (System A\footnote{System A: Supercell 
with 71 carbon atoms and 1 atomic vacancy.})  & 19.4\\    
Graphene sheet with one single vacancy\\ (System B\footnote{System B: Supercell 
with 161 carbon atoms and 1 atomic vacancy.})  & 18.8\\
Graphene sheet with two single vacancies\\(System C\footnote{System C: 
Supercell 
with 160 carbon atoms and 2 atomic vacancies ca. 11 Å apart.}) & 18.3\\
Graphene sheet with two single vacancies \\(System D\footnote{System D: 
Supercell 
with 160 carbon atoms and 2 atomic vacancies ca. 4.5 Å apart.})& 20.2\\
Graphite nanoribbon \\(System E\footnote{System E: Supercell with 96 carbon 
atoms 
with zigzag edges partially passivated by 16 oxygen atoms.}) & 20.8\\
\hline
\end{tabular}
\end{center}
\end{table}

There were some variations in the average B$_{hf}$ values as well as in the 
total 
magnetic moments of each system due to the possible interactions between 
neighbor atomic magnetic moments. As an example, the comparison between systems 
containing a single atomic vacancy (systems A and B, as described in 
Table \ref{systems}) showed a slightly reduced B$_{hf}$ value for the larger 
supercell, corresponding to the larger separation between each defect and its 
image produced by the use of periodic boundary conditions in the DFT 
calculations. Similarly, in the case of two interacting single vacancies 
(systems C and D), the B$_{hf}$ and the total magnetic moment were found to 
increase 
with the reduction in the separation between the two ferromagnetically-coupled 
vacancies. The largest total magnetic moments and B$_{hf}$ values among the 
studied 
systems were found for the graphite nanoribbon, where the magnetic response is 
the result of the ferromagnetic coupling between the magnetic moments 
associated with dangling bonds at edge sites \cite{silva}.

\begin{figure}[h!]
\begin{center}
\includegraphics[width= 8.5cm]{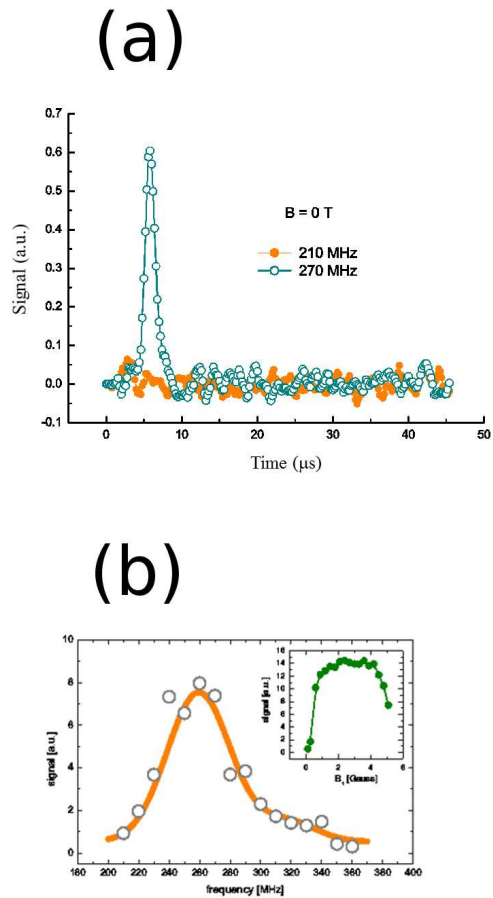}
\caption{(a) Raw time-domain zero-field NMR signal due to $^{13}$ nuclei in 
ferromagnetic graphite at two frequencies (full symbols 210 MHz, empty symbols 
270 MHz). The spin echo is clearly visible at 270 MHz, whereas the 'blank' 
measurement at 210 MHz demonstrates the absence of any spurious baseline 
signals. This zero-field NMR signal is strong evidence of a large and intrinsic 
hyperfine magnetic field at the carbon sites. b) $^{13}$ zero-field NMR 
spectrum, obtained by integrating spin echoes at different frequencies. The 
measurement was performed at 1.5 K. Two distinct NMR lines are resolved, the 
faint upper (U) and much stronger lower (L) line (the solid line is a double 
gaussian fit). Inset shows the dependence of signal intensity on the excitation 
pulse strength, with a shape characteristic of ferromagnetic materials.}
\label{fig2}
\end{center}
\end{figure}

Guided by these DFT calculations, zero-field NMR experiments were then carried 
out in a frequency range encompassing the resonance frequencies corresponding 
to the predicted B$_{hf}$ values. Using the well-known magnetogyric ratio of 
$^{13}$C (10.569 MHz/T) \cite{harris}, these frequencies were estimated in 
the range 200-230 MHz. A significant zero-field NMR signal was indeed found at 
frequencies just above this range at 1.5 K (Fig.\ref{fig2}a); by sweeping the 
frequency, the maximum signal was detected at a frequency of ca. 260 MHz, as 
shown in the NMR spectrum exhibited in Fig.\ref{fig2}b. The B$_{hf}$ value 
corresponding to this peak is around 24 T; the deviation from the calculated 
values given in Table 1 is not surprising, considering the complexity of the 
real material (from the chemical, structural and magnetic points of view) in 
comparison to the idealized model systems used in the DFT calculations.
The NMR spectrum shown in Fig. \ref{fig2} was fitted by using two Gaussian 
lines, named lower (L) and upper (U) lines, with different intensities (L being 
thus the dominant contribution in zero-field NMR spectra). The dependence of L 
peak intensity on RF pulse field amplitude B1 (Fig. \ref{fig2}b, inset) clearly 
did not follow the usual sinusoidal curve observed in diamagnetic materials: 
this is characteristic of NMR in ferromagnets and is caused by B1 and signal 
enhancement due to the oscillation of large electronic magnetic moments 
specially in domain walls \cite{turov}. A $\pi$/2 pulse in the regular sense 
is thus not well defined, making absolute intensity comparison between different 
lines somewhat ambiguous; this problem was exacerbated in NMR experiments 
performed under external magnetic fields, as discussed below. However, the shape 
of the echo intensity versus B1 curve is strong evidence that the observed 
spectrum was indeed associated with the ferromagnetic-enhanced response due to 
$^{13}$ nuclei in ferromagnetic graphite. It is worth noting that the features 
of the NMR spectrum shown in Fig. \ref{fig2} were well reproduced in an 
independent experiment performed with a second ferromagnetic graphite sample 
from another batch, confirming the robustness of the method of sample 
preparation and the intrinsic character of the zero-field NMR signal detected in 
this material.

\begin{figure}[h!]
\begin{center}
\includegraphics[width= 8.5cm]{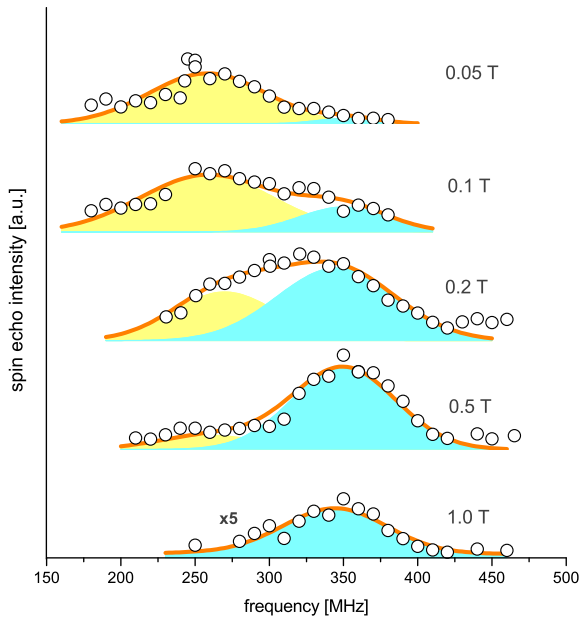}
\caption{C NMR spectra recorded under several external magnetic fields for 
ferromagnetic graphite, at 1.5 K. The spectrum at 1 T is magnified five times 
for clarity. Solid lines are double Gaussian fits. The two NMR lines observed in 
zero field show rather different behavior in applied fields, indicating a 
difference in their origin.}
\label{fig3}
\end{center}
\end{figure}

The relative intensities of the U and L lines were found to change dramatically 
with the application of small external magnetic fields (Fig. \ref{fig3}), 
suggesting a different origin for the L and U signals. The monotonous drop of 
the L signal intensity suggests that the L line originates from nuclei in Bloch 
walls, since the number of domains is expected to decrease with increasing 
external field. Following this simple reasoning, the U line would be due to 
$^{13}$ nuclei inside ferromagnetic domains. Accordingly, the peak position of 
this line was found to shift by ca. –10 MHz in an external field of 1 T, 
following the shift expected based on the magnetogyric ratio of $^{13}$C nuclei 
for the case where the external magnetic field is collinear with the hyperfine 
magnetic field.

In summary, $^{13}$C NMR experiments performed under zero or small applied 
magnetic fields in conjunction with DFT calculations carried out in model 
systems allowed the direct evaluation of the hyperfine magnetic field in a 
ferromagnetic carbon-based material, opening the possibility of more in-depth 
studies of the hyperfine interaction in carbon materials presenting magnetic 
properties.

\begin{acknowledgements}
 The authors thank Damir Pajić (from the Laboratory for magnetic measurements at 
the University of Zagreb) for the SQUID magnetization measurements. JCCF and WLS 
are grateful to Prof. Tito Bonagamba (from the Physics Institute of São Carlos, 
University of São Paulo, Brazil) for his help with the development of the 
computational infrastructure used in this work. The financial support from the 
agencies FAPES, FAPESP, FINEP, CAPES and CNPq (Brazil) and HRZZ (Croatia) is 
also gratefully acknowledged.
\end{acknowledgements}

\section*{references}


\begin{thebibliography}{10}
\bibitem{request}. Information on materials and methods are available upon 
request.
\end{thebibliography}
\end{document}